\documentclass[12pt]{amsart}
\usepackage{graphicx,amsmath,amssymb}

\newcommand{\C}{\mathbb C}
\newcommand{\R}{\mathbb R}
\newcommand{\Z}{\mathbb Z}
\newcommand{\N}{\mathbb N}
\renewcommand{\d}{\prime}
\newcommand{\dd}{{\prime \prime}}

\newtheorem{theorem}{Theorem}
\newtheorem{lemma}[theorem]{Lemma}

\newtheorem{corollary}[theorem]{Corollary}

\newtheorem*{remark}{Remark}
\newtheorem*{remarks}{Remarks}

\begin{document}
\title[]
{On Half-Line Spectra for a Class of \\ Non-Self-Adjoint Hill Operators}
\author[]
{Kwang C. Shin}
\address{Department of Mathematics , University of Missouri, Columbia, MO
65211, USA}
\date{August 11, 2003}

\begin{abstract}
In 1980, Gasymov showed that non-self-adjoint Hill operators with
complex-valued periodic  potentials of the type $V(x)=\sum_{k=1}^{\infty}a_k
e^{i k x}$, with $\sum_{k=1}^{\infty}|a_k|<\infty$, have spectra
$[0,\,\infty)$. In this note, we provide an alternative and elementary proof
of this result.
\end{abstract}

\maketitle

\baselineskip = 18pt
%
\section{Introduction}

We study the Schr\"odinger equation
\begin{equation}\label{main_eq}
-\psi^\dd(z,x)+V(x)\psi(z,x)=z\psi(z,x),\quad x\in\R,
\end{equation}
where $z\in\C$ and $V\in L^{\infty}(\R)$ is a continuous complex-valued
periodic function of  period $2\pi$, that is, $V(x+2\pi)=V(x)$ for all
$x\in\R$.  The {\it Hill operator} $H$ in $L^2(\R)$ associated with
(\ref{main_eq}) is defined by
$$(Hf)(x)=-f^\dd(x)+V(x)f(x),\quad f\in W^{2,2}(\R),$$
where $W^{2,2}(\R)$ denotes the usual Sobolev space.
Then $H$ is a densely defined closed operator in $L^2(\R)$ (see, e.g.,
\cite[Chap. 5]{MSPE}).

The spectrum of $H$ is purely continuous and a union of countablely
many analytic arcs in the complex plane \cite{FSRB}. In general it is not easy to explicitly
determine the spectrum of $H$ with specific potentials. However, in 1980,
Gasymov \cite{Gasy1} proved the following remarkable result:
\begin{theorem} [\cite{Gasy1}] \label{main_thm}
Let $V(x)=\sum_{k=1}^{\infty}a_ke^{ikx}$ with
$\{a_k\}_{k\in\N}\in\ell^1(\N)$. Then the spectrum of the  associated Hill
operator $H$ is purely continuous and equals
$[0,\,\infty)$.
\end{theorem}
In this note we provide an alternative and elementary proof of this result.
Gasymov \cite{Gasy1} proved the existence of a solution $\psi$
of (\ref{main_eq}) of the form
$$
\psi(z,x)=e^{i\sqrt{z}x}\left(1+\sum_{j=1}^{\infty}\frac{1}{j+2\sqrt{z}}
\sum_{k=j}^{\infty}\nu_{j,k}e^{ik x}\right),
$$
where the series
$$\sum_{j=1}^{\infty}\frac{1}{j}\sum_{k=j+1}^{\infty}k(k-j)
|\nu_{j,k}|\,\,\text{ and }\,\,\sum_{j=1}^{\infty}j|\nu_{j,k}|$$
converge, and used this fact to show that the spectrum of $H$ equals
$[0,\,\infty)$. He also discussed the corresponding inverse spectral
problem. This inverse spectral problem was subsequently considered by
Pastur and Tkachenko \cite{Pastur} for $2\pi$-periodic potentials in
$L^2_{{\rm loc}}(\R)$ of the form $\sum_{k=1}^{\infty}a_ke^{ikx}$.

In this paper, we will provide an elementary proof of the following result.
\begin{theorem}\label{main_thm2}
Let $V(x)=\sum_{k=1}^{\infty}a_ke^{ikx}$ with
$\{a_k\}_{k\in\N}\in \ell^1(\N)$. Then
\begin{equation*}
\Delta(V;z)=2\cos(2\pi\sqrt{z}),
\end{equation*}
where $\Delta(V;z)$ denotes the Floquet discriminant associated with
\eqref{main_eq} (cf.\ equation \eqref{Del}).
\end{theorem}

\begin{corollary}\label{COR}
 Theorem \ref{main_thm2} 
implies that the spectrum of $H$ equals $[0,\,\infty)$; it also implies Theorem \ref{main_thm}.
\end{corollary}
\begin{proof}
In general, one-dimensional Schr\"odinger operators  with periodic
potentials have purely continuous spectra (cf.\ \cite{FSRB}). Since $-2\leq 2\cos(2\pi\sqrt{z})\leq 2$ if and only if
$z\in[0,\,\infty)$, one concludes that the spectrum of $H$ equals $[0,\,\infty)$ and that Theorem \ref{main_thm} holds (see Lemma \ref{lemma_equiv} below).
\end{proof}

\begin{remark}
{\rm We note that the potentials $V$ in Theorem \ref{main_thm} are nonreal
and hence $H$ is non-self-adjoint in $L^2(\R)$ except when $V=0$. It is known
that $V=0$ is the only {\it real} periodic potential for which the spectrum
of
$H$ equals $[0,\,\infty)$ (see \cite{BORG}). However, if we allow the
potential $V$ to be complex-valued, Theorem \ref{main_thm} provides
a family of complex-valued potentials such that spectra of the associated
Hill operators equal $[0,\,\infty)$. From the point of view of inverse
spectral theory this yields an interesting and significant nonuniqueness
property of non-self-adjoint Hill operators in stark contrast to self-adjoint
ones. For an explanation of this  nonuniqueness property of non-self-adjoint
Hill operators in terms of associated Dirichlet eigenvalues, we refer to
\cite[p.\ 113]{GH03}.}
\end{remark}

As a final remark we mention some related work of
Guillemin and Uribe \cite{Guill2}.
Consider the differential equation (\ref{main_eq}) on the interval
$[0,\,2\pi]$ with the periodic boundary conditions. It is shown in
\cite{Guill2} that all potentials in Theorem \ref{main_thm}
generate the same spectrum $\{n^2: n=0, 1, 2,\ldots\}$, that is,
$\Delta(V;n^2)=2$  for all $n=0, 1, 2, \ldots.$

\section{Some known facts}\label{sec2}
In this section we recall some definitions and known results regarding
(\ref{main_eq}).  

For each $z\in\C$, there exists a fundamental system of
solutions
$c(V;z,\,x),\, s(V;z,\,x)$ of (\ref{main_eq}) such that
\begin{eqnarray}
c(V;z,\,0)=1,&& c^\d(V;z,\,0)=0, \nonumber\\
s(V;z,\,0)=0,&& s^\d(V;z,\,0)=1,\nonumber
\end{eqnarray}
where we use ${}^\d$ for $\frac{\partial}{\partial x}$ throughout this note.
The {\it Floquet discriminant} $\Delta(V;z)$  of (\ref{main_eq}) is
then defined by
  \begin{equation}
\Delta(V;z)=c(V;z,\,2\pi)+s^\d(V;z,\,2\pi). \label{Del}
\end{equation}
The Floquet discriminant $\Delta(V;z)$ is an  entire function
of order $\frac{1}{2}$ with respect to $z$ (see \cite[Chap. 21]{ECT}).

\begin{lemma}\label{exist_thm}
For every $z\in\C$ there exists a solution $\psi(z,\cdot)\not=0$ of
(\ref{main_eq}) and a number $\rho(z)\in\C\setminus\{0\}$ such that
$\psi(z,x+2\pi)=\rho(z)\psi(z,x)$ for all $x\in\R$. Moreover,
\begin{equation}\label{frac_eq}
\Delta(V;z)=\rho(z)+\frac{1}{\rho(z)}.
\end{equation}
In particular, if $V=0$, then $\Delta(0;z)=2\cos(2\pi\sqrt{z})$.
\end{lemma}

For obvious reasons one calls $\rho(z)$ the {\it Floquet
multiplier} of equation (\ref{main_eq}).

\begin{lemma}\label{lemma_equiv}
Let $H$ be the Hill operator associated with (\ref{main_eq}) and $z\in\C$.
Then the following four assertions are equivalent:
\begin{enumerate}
\item [(i)] $z$ lies in the spectrum of $H$.
\item [(ii)] $\Delta(V;z)$ is real and $|\Delta(V;z)|\leq 2$.
\item [(iii)] $\rho(z)=e^{i\alpha}$ for some $\alpha\in\R$.
\item [(iv)] Equation (\ref{main_eq}) has a
non-trivial bounded solution $\psi(z,\cdot)$ on $\R$.
\end{enumerate}
\end{lemma}
For proofs of Lemmas \ref{exist_thm} and \ref{lemma_equiv}, see, for
instance, \cite[Chs.\ 1, 2, 5]{MSPE}, \cite{MW}, \cite{FSRB} (we note that
$V$ is permitted to be locally integrable on $\R$).

\section{Proof of Theorem \ref{main_thm2}}
In this section we prove Theorem \ref{main_thm2}. In doing so, we
will use the standard identity theorem in complex analysis which asserts
that two analytic functions coinciding on an infinite set with an
accumulation point in their common domain of analyticity, in fact coincide
throughout that domain. Since both $\Delta(V;z)$ and $2\cos(2\pi\sqrt{z})$
are entire functions, to prove that $\Delta(V;z)=2\cos(2\pi\sqrt{z})$, it
thus suffices to  show that $\Delta(V;1/n^2)=2\cos(2\pi/n)$ for all integers
$n\geq 3$.

Let $n\in\N$, $n\geq 3$ be fixed and let $\psi\not=0$ be the solution of
(\ref{main_eq}) such that $\psi(z,x+2\pi)=\rho(z)\psi(z,x)$, $x\in\R$
for some $\rho(z)\in\C$. The existence of such $\psi$ and
$\rho$ is guaranteed by Lemma \ref{exist_thm}. We set
$\phi(z,x)=\psi(z,nx)$. Then
\begin{equation}\label{exp_eq}
\phi(z,x+2\pi)=\rho^n(z)\phi(z,x),
\end{equation}
and
\begin{equation}\label{eq_1}
-\phi^\dd(z,x)+q_n(x)\phi(z,x)=n^2z\phi(z,x),
\end{equation}
where
\begin{equation}\label{q_eq}
q_n(x)=n^2V(nx)=n^2\sum_{k=1}^{\infty}a_ke^{iknx},
\end{equation}
with period $2\pi$.
  Moreover, by (\ref{frac_eq}) and (\ref{exp_eq}),
\begin{equation}\label{rel_eq}
\Delta(q_n;w)=\rho^n(z)+\frac{1}{\rho^n(z)},\,\text{ where }\, w=n^2z.
\end{equation}

We will show below that
\begin{equation}\label{iden_eq}
\Delta(q_n;1)=2\,\text{ for every positive integer } n\geq 3.
\end{equation}
First, if $w=1$ (i.e., if $z=\frac{1}{n^2}$), then the fundamental
system of solutions $c(q_n;1,x)$ and $s(q_n;1,x)$ of (\ref{eq_1}) satisfies
\begin{eqnarray}
c(q_n;1,x)&=&\cos(x)+\int_0^x\sin(x-t)q_n(t)c(q_n;1,t)\,dt,\label{int_eq1}\\
s(q_n;1,x)&=&\sin(x)+\int_0^x\sin(x-t)q_n(t)s(q_n;1,t)\,dt.\nonumber
\end{eqnarray}
Moreover, we have
\begin{equation}\label{der_eq}
s^\d(q_n;1,x)=\cos(x)+\int_0^x\cos(x-t)q_n(t)s(q_n;1,t)\,dt.
\end{equation}

We use the Picard iterative method of solving the above integral equations.
Define sequences $\{u_j(x)\}_{j\geq 0}$ and $\{v_j(x)\}_{j\geq 0}$ of
functions as follows.
\begin{align}
u_0(x)=\cos(x),\quad&u_j(x)=\int_0^x\sin(x-t)q_n(t)u_{j-1}(t)\,dt,
\label{rec_eq1}\\
v_0(x)=\sin(x),\quad&v_j(x)=\int_0^x\sin(x-t)q_n(t)v_{j-1}(t)\,dt,
\quad j\geq 1.\label{rec_eq2}
\end{align}
Then one verifies in a standard manner that
\begin{equation}\label{eq_3}
c(q_n;1,x)=\sum_{j=0}^{\infty}u_j(x),\quad s(q_n;1,x)
=\sum_{j=0}^{\infty}v_j(x),
\end{equation}
where the sums converge uniformly over $[0,2\pi]$.
Since
\begin{equation*}
\Delta(q_n;1)=c(q_n;1,2\pi)+s^\d(q_n;1,2\pi),
\end{equation*}
to prove that $\Delta(q_n;1)=2$, it suffices to show that
the integrals in (\ref{int_eq1}) and (\ref{der_eq}) vanish at $x=2\pi$.

Next, we will rewrite (\ref{rec_eq1}) as
\begin{eqnarray}
u_0(x)&=&\frac{1}{2}(e^{ix}+e^{-ix}),\nonumber\\
u_j(x)&=&\frac{e^{ix}}{2i}\int_0^x e^{-it}q_n(t)u_{j-1}(t)\,dt
-\frac{e^{-ix}}{2i}\int_0^x e^{it}q_n(t)u_{j-1}(t)\,dt,\quad
j\geq 1.\label{rec_eqs}
\end{eqnarray}
Using this and (\ref{q_eq}), one shows by induction
on $j$ that $u_j$, $j\geq 0$, is of the form
\begin{equation}\label{smpow}
u_j(x)=\sum_{\ell=-1}^{\infty}b_{j,\ell}\,e^{i\ell x} \,
\text{ for some }\, b_{j,\ell}\in\C,
\end{equation}
the sum converging uniformly for $x\in\R$.  This follows from $n\geq 3$
because the smallest exponent of $e^{it}$ that $q_n u_{j-1}$ can have in
(\ref{rec_eqs}) equals $2$. (The first three terms in (\ref{smpow}) are due
to the antiderivatives of $e^{\pm it}q_n(t)u_{j-1}(t)$, evaluated at $t=0$.)
Next we will use  (\ref{eq_3}) and (\ref{smpow}) to show that
\begin{equation}\label{eq_eva}
\int_0^{2\pi}\sin(2\pi-t)q_n(t)c(q_n;1,t)\,dt=0.
\end{equation}

We begin with
\begin{eqnarray}
& &\int_0^{2\pi}\sin(2\pi-t)q_n(t)c(q_n;1,t)\,dt\nonumber\\
&=&-\frac{1}{2i}\int_0^{2\pi}(e^{it}-e^{-it})q_n(t)c(q_n;1,t)\,dt
\nonumber\\
&=&-\frac{1}{2i}\int_0^{2\pi}(e^{it}-e^{-it})\left(\sum_{k=1}^{\infty}
a_ke^{iknt}\right)\left(\sum_{j=0}^{\infty}u_j(t)\right)\,dt\nonumber\\
&=&-\frac{1}{2i}\sum_{k=1}^{\infty}\sum_{j=0}^{\infty}a_k\int_0^{2\pi}
(e^{i(kn+1)t}-e^{i(kn-1)t})u_j(t)\,dt\label{integral_ex},
\end{eqnarray}
where the change of the order of integration and summations is permitted due
to the uniform convergence of the sums involved.
The function $(e^{i(kn+1)t}-e^{i(kn-1)t})u_j(t)$ is a power series
in $e^{it}$ with no constant term (cf.\ (\ref{smpow})), and hence its
antiderivative is a periodic function of period $2\pi$. 
Thus, every integral in (\ref{integral_ex}) vanishes, and hence
(\ref{eq_eva}) holds.  So from (\ref{int_eq1}) we conclude that
$c(q_n;1,2\pi)=1$.

Similarly, one can show by induction that $v_j$ for each $j\geq 0$
is of the form (\ref{smpow}). Hence, from (\ref{der_eq}), one concludes that
$s^\d(q_n;1,2\pi)=1$ in close analogy to the proof of $c(q_n;1,2\pi)=1$.
Thus, (\ref{iden_eq}) holds for each $n\geq 3$.

So by (\ref{rel_eq}),
\begin{equation*}
\Delta(q_n;1)=\rho^n(1/n^2)+\frac{1}{\rho^n(1/n^2)}=2 \,
\text{ for every }\,n\geq 3.
\end{equation*}
This implies that $\rho^n(1/n^2)=1$. So
$\rho(1/n^2)\in\{\xi\in\C:\xi^n=1\}$. Thus,
$\Delta(V;1/n^2)\in\{2\cos(2k\pi/n):k\in\Z\}$. Next, we will
show that
$\Delta(V;1/n^2)=2\cos(2\pi/n)$.

We consider a family of potentials $q_{\varepsilon}(x)=
\varepsilon V(x)$ for $0\leq \varepsilon\leq 1$. For each
$0\leq\varepsilon\leq 1$, we apply the above argument to get that
$\rho(\varepsilon,1/n^2)\in\{\xi\in\C:\xi^n=1\}$, where we use the notation
$\rho(\varepsilon,1/n^2)$ to indicate the possible
$\varepsilon$-dependence of $\rho(1/n^2)$. Next, by the integral
equations  (\ref{int_eq1})--(\ref{rec_eq2}) with
$q_{\varepsilon}={\varepsilon}V$ instead of $q_n$, one sees that
$\Delta(\varepsilon V;1/n^2)$ can be written as a power series
in $\varepsilon$ that converges uniformly for
$0\leq \varepsilon\leq 1$.
Thus, the function $\varepsilon\mapsto\Delta(\varepsilon V;1/n^2)\in
\{2\cos(2k\pi/n):k\in\Z\}$ is continuous for
$0\leq \varepsilon\leq 1$ (in fact, it is entire w.r.t. $\varepsilon$).  Since
$\{2\cos(2k\pi/n):k\in\Z\}$ is discrete, and since
$\Delta(\varepsilon V;1/n^2)=2\cos(2\pi/n)$ for $\varepsilon=0$,
we conclude
that
$$
\Delta(\varepsilon V;1/n^2)=\Delta(0;1/n^2)=2\cos (2\pi/n)
\, \text{ for all }\,0\leq\varepsilon\leq 1.$$
In particular, $\Delta(V;1/n^2)=2\cos (2\pi/n)$ for every positive integer
$n\geq 3$. Since $\Delta(V;z)$ and $2\cos(2\pi\sqrt{z})$ are both
entire and since they coincide at $z=1/n^2$, $n\geq 3$, we conclude that
$$\Delta(V;z)=2\cos(2\pi\sqrt{z})\,\text{ for all }\,z\in\C$$
by the identity theorem for analytic functions alluded to at the beginning
of this section. This completes proof of Theorem \ref{main_thm2} and hence
that of  Theorem \ref{main_thm} by Corollary \ref{COR}.

\begin{remarks}
{\rm (i) Adding a constant term to the potential $V$ yields a translation
of the spectrum. (ii) If the potential $V$ is a power series in $e^{-ix}$
with no constant term, then the spectrum of $H$ is still
$[0,\,\infty)$, by complex conjugation. (iii) If $V$ lies in the $L^2([0,\,2\pi])$-span of
$\{e^{ikx}\}_{k\in\N}$, then by continuity of
$V\mapsto\Delta(V;z)$ one concludes $\Delta(V;z)=2\cos(2\pi\sqrt{z})$
and hence the spectrum of $H$ equals $[0,\,\infty)$ (see
\cite{Pastur}). \\
(iv) Potentials $V$ that include negative and positive integer powers
of $e^{ix}$  are not included in our note. Consider, for example, equation
(\ref{main_eq}) with $V(x)=2\cos(x)$, the so-called Mathieu equation. The
spectrum of $H$ in this case is known to be a  disjoint union of
infinitely many closed intervals on the real line \cite{INC} (also, see
\cite{MSPE}, \cite{MW}). In particular,  the spectrum of $H$ is not
$[0,\,\infty)$.  In such a case the antiderivatives of
$(e^{i(kn+1)t}-e^{i(kn-1)t})u_j(t)$ in  (\ref{integral_ex}) need not be
periodic and our proof breaks down. }
\end{remarks}
\noindent {\bf Acknowledgments.} The author thanks Fritz Gesztesy and
Richard Laugesen for suggestions  and discussions to improve the presentation of this note.

{\sc e-mail:}  kcshin@math.missouri.edu
\end{document}